\documentclass[nofootinbib,article,
pre,
twocolumn,
eqsecnum,
amsmath,amssymb,
amsmath,amssymb, aps,
]{revtex4}
\usepackage[dvipsnames]{xcolor} 
\usepackage{epsfig}
\usepackage{amsfonts}
\usepackage{amsmath}
\usepackage{wasysym}
\usepackage{amssymb}
\usepackage{bm} 
\usepackage[framemethod=PStricks]{mdframed}
\usepackage{lipsum}
\usepackage{float}

\usepackage{pdflscape}
\usepackage{graphicx}
\usepackage{ulem}
\usepackage{sidecap}

\newcommand{\be}{\begin{equation}}
\newcommand{\ee}{\end{equation}} 
\newcommand{\bea}{\begin{eqnarray}} 
\newcommand{\eea}{\end{eqnarray}}
\newcommand{\bfr}{{\bf{r}}}

\usepackage[utf8]{inputenc}
\usepackage{bbm} 				
\usepackage{epstopdf}				
\usepackage{verbatim}    			
\usepackage{sidecap}                
\usepackage{indentfirst}
\usepackage[colorinlistoftodos]{todonotes}

\begin{document}

\title{Statistics of Extreme Turbulent Circulation \\ Events
from Multifractality Breaking}
\author{L. Moriconi$^{1,}$\footnote{Corresponding author: moriconi@if.ufrj.br} and R.M. Pereira$^{2}$}
\affiliation{$^{1}$Instituto de F\'\i sica, Universidade Federal do Rio de Janeiro, \\
C.P. 68528, CEP: 21945-970, Rio de Janeiro, RJ, Brazil}
\affiliation{$^{2}$Instituto de Física, Universidade Federal Fluminense, 24210-346 Niterói, RJ, Brazil}


\begin{abstract}
Recent numerical explorations of extremely intense circulation fluctuations at high Reynolds number flows have brought to light novel aspects of turbulent intermittency. {\textcolor{black}{Vortex gas modeling ideas, which are related to a picture of turbulence as a dilute system of vortex tube structures}}, have been introduced alongside such developments, leading to accurate descriptions of the core and the intermediate tails of circulation probability distribution functions (cPDFs), as well as the scaling exponents associated to statistical moments of circulation. We extend the predictive reach of the vortex gas picture of turbulence, by emphasizing that multifractality breaking, one of its salient phenomenological ingredients,
is the key concept to disclose the asymptotic form of cPDF tails. 
A remarkable analytical agreement is found with previous results derived within 
the framework of the instanton approach to circulation intermittency, {\textcolor{black}{a functional formalism devised to single out the statistically dominant velocity configurations associated to extreme circulation events.}}
\end{abstract}


\maketitle


\section{Introduction}

Homogeneous and isotropic turbulent flows are usually associated to the intensification and mixing of the vorticity field across a broad range of length scales \cite{tenn-lum}. Compelling evidence has been gathered since the mid-1990's, as the result of direct numerical simulations (DNS), to reveal that the flow regions where vorticity is the most intense are shaped like elongated thin vortex tubes \cite{she_etal,farge_etal,yokokawa_etal,kaneda_etal}. These are entangled and strongly interacting long-lived coherent structures, which account for essentially all of the turbulent kinetic energy that cascades down from the largest to the smallest scales of dynamical importance \cite{farge_etal}.

Relying on the above phenomenological hints, one might expect velocity circulation to be a key observable in the statistical description of turbulence. Around three decades ago, actually, Migdal imported to the context of fluid dynamics high-energy functional methods \cite{migdal1994} to study the far, supposedly non-gaussian, tails of circulation probability distribution functions (cPDFs). However, the first subsequent investigations carried out within the numerical and experimental fronts \cite{umeki,cao_etal,benzi_etal1} were unfortunately unable to trigger continued progress, mainly due to the existing computational limitations of the time (both in speed and memory capacity). 

More recently, considerable hardware improvements have enabled the implementation of DNS at much higher Reynolds numbers, so that a vivid interest in the problem of turbulent circulation statistics has resurfaced in the literature \cite{Iyer_etal,apol_etal,migdal2020,mori1,Iyer_etal_PNAS,mori_PNAS,mori_pereira_valadao}, even driving further perspectives in the understanding of quantum turbulence \cite{muller_etal,polanco_etal}. The deadlock of numerical issues was broken by a computational tour de force performed by Iyer et al. \cite{Iyer_etal,Iyer_etal_PNAS}, who have identified a number of relevant statistical aspects of the turbulent circulation $\Gamma$, summarized as follows: 

\noindent (i) Scaling exponents of its statistical moments depart more clearly from the Kolmogorovian-like predicted values at high orders, where they become linearly dependent on the moment orders; 

\noindent (ii) Circulation kurtoses grow with non-elementary functional forms as length scales get smaller, saturating, at the bottom of the inertial range, with weakly Reynolds number dependent values; 

\noindent (iii) cPDFs' far tails have simple exponential forms modulated by prefactors $\propto 1/\sqrt{|\Gamma|}$; 

\noindent (iv) Properly rescaled cPDFs' tails collapse for contours that span identical minimal surface areas.

A recent vortex gas model, which combines the multiplicative cascade nature of turbulence and its structural elements ({\textcolor{black}{that is, a dilute set of}} vortex tubes taken as the spatial support of relevant dynamical degrees of freedom) has led to {\textcolor{black}{quantitative accounts}} of the above items (i) and (ii) \cite{apol_etal,mori1,mori_pereira_valadao}, {\textcolor{black}{which accurately compare to the findings of Ref.~\cite{Iyer_etal}}}.

We aim in this work to take a deeper look into the foundations of the vortex gas model and to show that property (iii) {\textcolor{black}{\cite{migdal2020,Iyer_etal_PNAS}}}, a most distinctive signature of circulation intermittency, can also be recovered along the same modeling guidelines. We will not touch on point (iv), waved here only for the sake of information completeness. That means, in practical terms, that we restrict our analysis to the circulation evaluated on oriented contours which enclose planar domains.
\vspace{-0.0cm}

This paper is organized as follows. Sec.~II outlines the principal ideas of the vortex gas model, which are additionally strengthened in Secs. III and IV, through the analysis of DNS data and Monte Carlo simulations. Assembling the modeling ingredients then collected and working with simulational-assisted arguments, we derive, in Sec.~V, the functional form of asymptotic cPDF tails, which is surprisingly noted to agree with (completely independent) results derived from previous analytical studies. In Sec.~VI, finally, we summarize our findings and point out directions of further research.

\section{Modeling Principles}

The main motivation underlying the vortex gas approach to circulation statistics \cite{apol_etal} can be synthetized in Fig.~1. The image there depicted is produced from a DNS of homogeneous and isotropic turbulence with Taylor-Reynolds number $R_\lambda = 433$, as available from the Johns Hopkins turbulent database \cite{JHTD}. The picture shows, through an application of the swirling strength criterion for the identification of vortex structures \cite{zhou_etal}, a large number of disconnected spots which are assumed to locate the intersections of {\textcolor{black}{putative}} three-dimensional vortex tubes \cite{she_etal,farge_etal,yokokawa_etal,kaneda_etal} with an arbitrary planar slice of the DNS domain.

{\textcolor{black}{The swirling strength methodology defines a point $\bfr$ in a slicing plane similar to the one of Fig.~1 to belong to a vortex structure if the planar velocity gradient tensor at $\bfr$ has a non-degenerate pair of complex conjugate eigenvalues $(\lambda , \bar \lambda)$, which indicates local swirling motions. Our procedure of cutting vortex tubes by ``tomographic" planes is actually a usual tool in the investigation of vortical coherent structures in wall-bounded turbulent flows \cite{zhou_etal,adrian_chris,herpin_etal,hugo_mori}.}}

The spots of Fig.~1 can be effectively taken, in this way, as planar vortex structures, which are found to have typical linear dimensions of a few times 
the Kolmogorov dissipation length $\eta_K$ \cite{mori_pereira_valadao}.

\begin{figure}[h]
\includegraphics[width=0.45\textwidth]{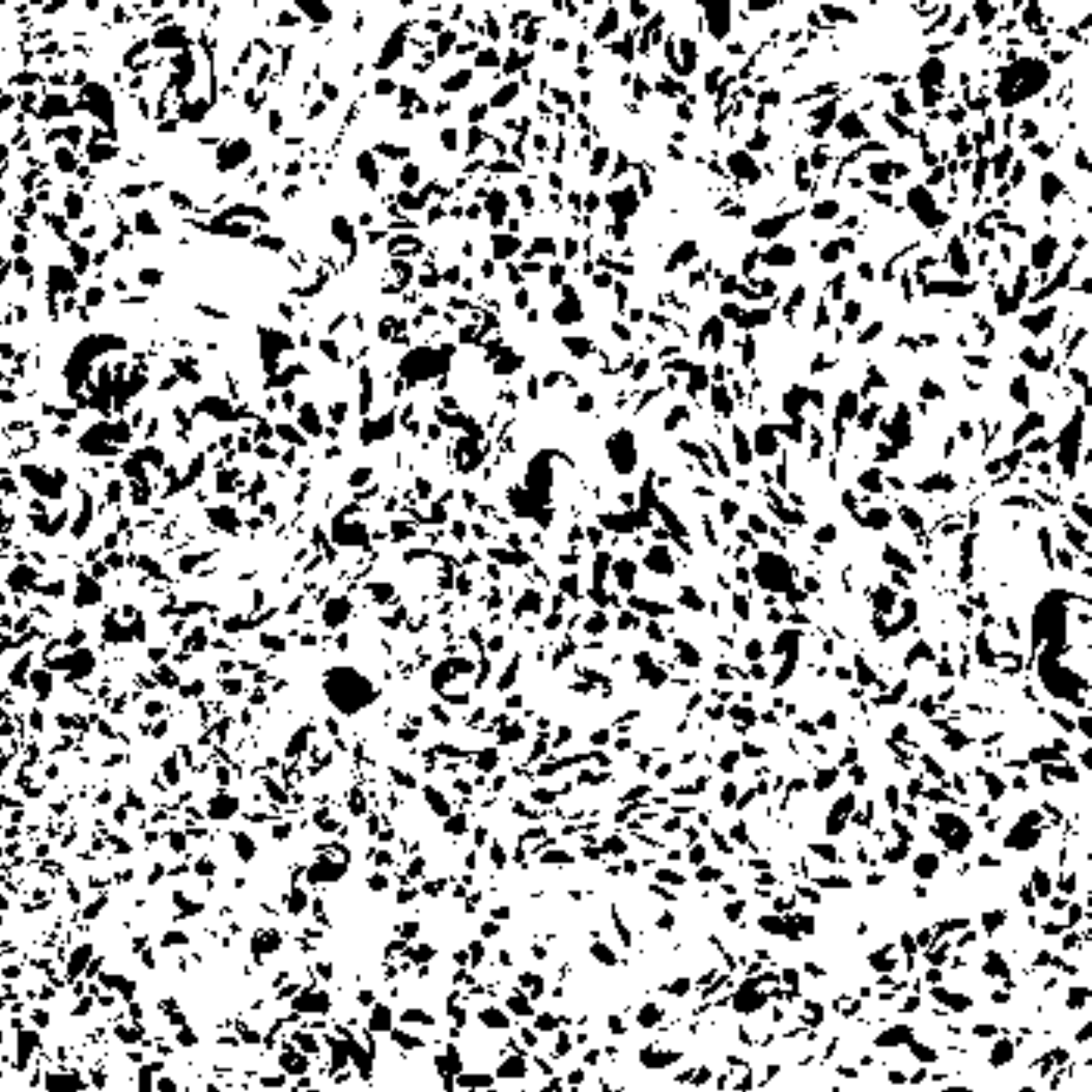}
\vspace{0.0cm}
\caption{\textcolor{black}{Planar vortex structures, represented as a set of disconnected dark spots, indicate the intersections of three-dimensional vortex tubes} with a $350 \times 350$ planar slice of a $1024^3$ DNS domain, as post-processed from the Johns Hopkins turbulence database \cite{JHTD}.}
\label{}
\end{figure}

We rephrase, throughout this section, the essential content of the vortex gas model and refer to \cite{apol_etal,mori1,mori_pereira_valadao} for {\textcolor{black}{supplementary technical details}}. 
Let then $\tilde \Gamma(\bfr_i)$ be the total circulation carried by a vortex spot centered at $\bfr_i$ and $\xi(\bfr_i)$ be the surface number density of vortex spots placed in some small planar neighborhood that surrounds $\bfr_i$. {\textcolor{black}{Working in a continuum approximation for the circulation around a planar domain $\mathcal{D}$ of area $A$, we have that}}
\be
\Gamma = \int_\mathcal{D} d^2 \bfr \,\xi(\bfr) \tilde \Gamma(\bfr)  \ , \ \label{circ} 
\ee
where, in the vortex gas model, $\xi(\bfr)$ is related, 
{\textcolor{black}{up to some unimportant dimensionless constant}}, to the energy dissipation rate field $\epsilon(\bfr)$ as 
{\textcolor{black}{
\be
\xi(\bfr) = \frac{1}{\eta_K^2} 
\sqrt{\frac{\epsilon(\bfr)}{\epsilon_0}} \ . \ \label{xi1}
\ee
Above, $\epsilon_0 \equiv \mathbb{E} [ \epsilon(\bfr) ]$
is the mean energy dissipation rate in the flow. 
Notice, however, that the statistical moments and the cPDF's of the standardized circulation $\Gamma / \sqrt{\mathbb{E} [ \Gamma^2 ]}$ do not depend on $\epsilon_0$, a fact that underlies many of the developments advanced in \cite{apol_etal,mori1,mori_pereira_valadao}.}}

The reduced circulation field $\tilde \Gamma(\bfr)$ is postulated to be a bounded random Gaussian field with vanishing mean and two-point correlation function
\be
{\mathbb{E}} [ \tilde \Gamma(\bfr) \tilde \Gamma(\bfr') ] \sim |\bfr - \bfr'|^{-\alpha} \label{gamma_c}
\ee
within inertial range scales, {\textcolor{black}{with $\alpha = 4/3 - \mu/2$}}, where $\mu$ is the intermittency scaling exponent that describes the spatial decay of energy dissipation rate correlations \cite{frisch}, that is, 
\be
{\mathbb{E}} [ \epsilon(\bfr) \epsilon(\bfr') ] \sim |\bfr - \bfr'|^{-\mu} \ . \ \label{epsilon_c}
\ee
{\textcolor{black}{The determination of the intermittency exponent $\mu$ has been the subject of some debate \cite{frisch}, with reasonable past estimates given by $\mu = 0.20 \pm 0.02$ \cite{antonia_etal} and
$\mu = 0.25 \pm 0.05$ \cite{sreeni_kaila}. We take here the value $\mu = 0.17 \pm 0.01$, recently established through the analysis of extensive experiments carried out with plane and circular jets \cite{tang_etal}}.

Resorting to the relatively slow decay (\ref{epsilon_c}), when compared to (\ref{gamma_c}), an alternative formulation of (\ref{circ})
has been put forward as
\be
\Gamma = \xi_{\hbox{cg}} (\mathcal{D}) \int_\mathcal{D} d^2 \bfr \tilde \Gamma(\bfr)  \ , \ \label{circ2} 
\ee
where
\be
\xi_{\hbox{cg}} (\mathcal{D}) \equiv \frac{1}{A} \int_\mathcal{D} d^2 \bfr \xi(\bfr) \label{cgxi}
\ee
is the coarse-grained version of $\xi(\bfr)$, taken over the domain $\mathcal{D}$. 
Eq. (\ref{circ2}) proves to be a very convenient tool for the statistical analysis of circulation fluctuations, as it follows from results of the field-theoretical generalization of the Obukhov-Kolmogorov (OK62) model of intermittency \cite{O62,K62}, known as the {\it{Gaussian multiplicative chaos}} (GMC) model of the turbulent cascade \cite{ro_va}. {\textcolor{black}{It can be shown, then, that (\ref{circ}) and (\ref{circ2}) behave, as a matter of fact, in the same scaling way with respect to domain size variations.}}

{\textcolor{black}{The GMC model of intermittency is a mathematical formalism which addresses in a rigorous way the postulated fundamental properties of the OK62 phenomenology. We mean, in particular, the lognormality of $\epsilon(\bfr)$ and the refined similarity hypothesis, viz., the statement that fluctuations of the three-dimensional coarse-grained dissipation rate scale as
\be
\mathbb{E} \left [ \left ( \ell^{-3} \int_{r \leq \ell} d^3 \bfr
\epsilon(\bfr) \right )^q \right ] \sim \ell^{\tau_q} \ , \ \label{GMC}
\ee
where $\tau_q = \mu q (1-q)/2$. An important heuristic point about the GMC modeling of homogeneous and isotropic turbulence is that it not only reproduces OK62, but it goes beyond: correlation functions of $\epsilon(\bfr)$ can be exactly computed, powers of $\epsilon(\bfr)$ are noticed to be lognormally distributed and to satisfy scaling relations similar to (\ref{GMC}), which, on its turn,
can be generalized to arbitrary subspaces of the three-dimensional space
(as we do for the two-dimensional slices of the flow).}}}

{\textcolor{black}{In short and concrete terms, the application of GMC modeling to a planar vortex gas like the one snapshotted in Fig.~1}} stands for representing the vortex surface density as the multifractal field
\be
\xi(\bfr) \propto \sqrt{\epsilon(\bfr}) \propto \exp [ \gamma \phi(\bfr) ] \ , \  \label{xi_gm}
\ee
where $\gamma = \sqrt{2 \pi \mu} \simeq 1.0$ and $\phi(\bfr)$ is a free massless two-dimensional scalar field \cite{zinn-justin}. It then follows, from GMC analysis, that $\xi_{\hbox{cg}}(\mathcal{D})$ is a lognormal random variable (which has its Reynolds number dependence discussed in \cite{apol_etal}). The product of (\ref{xi_gm}) with the usual two-dimensional integration measure $d^2 \bfr$ gives an example of what is called a {\it{Liouville measure}} in the lexicon of the GMC theory.

At this point, we note that a critical issue with the above modeling setup is that it leads to scaling exponents for the statistical moments of circulation that are quadratically dependent on their orders, failing to reproduce the clearly observed crossover from quadratic to linear behavior at high moment orders \cite{Iyer_etal}. It turns out, however, that a simple variation of the GMC basilar definitions yields a way to cope with this difficulty \cite{mori1}. The {\it{modified GMC model}} takes into account the phenomenon of multifractality breaking, translated here as the fact that fluctuations of $\xi(\bfr)$ are actually {\textcolor{black}{bounded from above}}. More specifically, while still keeping (\ref{xi_gm}), we now prescribe fluctuations of the scalar field $\phi(\bfr)$ to be ruled by the probability density functional $\exp \{ -S[\phi] \}/C$, 
where
$
C = \int D[\phi] \exp \{ -S[\phi] \}
$
and
\be
S[\phi] = \int_\mathcal{D} d^2 \bfr \left [  \frac{1}{2}  (\partial_i \phi)^2 + V(\phi) \right ] \ , \ \label{action}
\ee
with
\begin{equation}
V(\phi) =
    \begin{cases}
      0 \ , \ \text{if $\phi < \Phi_0 $} \ , \ \\
    V_0 \rightarrow \infty \ , \ \text{if $\phi \geq \Phi_0$} \ , \ \\
    \end{cases}    \label{V}
\end{equation}
for some phenomenologically chosen bounding parameter $\Phi_0$.
It is clear, in view of (\ref{xi_gm}) and (\ref{action}), that 
$\xi(\bfr)$ cannot fluctuate beyond $\exp [ \gamma \Phi_0 ]$.
{\textcolor{black}{The bound $\Phi_0$ is expected to have a logarithmic dependence upon $L/\eta_K$, where $L$ is the integral length scale of the flow \cite{mori1}. This implies that the upper-bound of $\xi(\bfr)$ scales with some power of the Reynolds number, but this is not going to be a point of relevance in our considerations.}}
As a trivial observation, we remark that the original GMC formulation can be refered to as the case ``$\Phi_0 = \infty$".

{\textcolor{black}{We also point out that the boundedness of $\xi(\bfr)$ (or $\epsilon(\bfr)$, equivalently) provides, as an important byproduct, a solution of a notorious difficulty of the OK62 model as a fully consistent description of intermittency. As emphasized by Frisch \cite{frisch}, the OK62 model does not respect the Novikov's convexity inequalities, which means that it fails to estimate scaling exponents of velocity structure functions at high enough moment orders. That would be the case here, if fluctuations of $\xi(\bfr)$ were not bounded. However, as it is already known from \cite{mori1}, bounded fluctuations of $\xi(\bfr)$ lead to scaling exponents of velocity structure functions which depart from the OK62 results at high orders, in sharp agreement with DNS evaluations \cite{Iyer_etal} and in consonance with the
Novikov's consistency prescriptions.}}

Before getting to the main arguments related to the analytical form of far cPDFs' tails, it is interesting (and important) to reassess, for validation purposes, the correctness of Eqs.~(\ref{xi1}) and (\ref{circ2}). This is our task in the next two sections. 

\section{Surface Vortex Density as a Multifractal Field}

{\textcolor{black}{Define the planar domain $\mathcal{D}_\ell$ to be a square of side $\ell$. 
According to the general results of the GMC theory about the statistical properties of the energy dissipation field \cite{ro_va}, 
it follows, from Eqs.~(\ref{xi1}) and (\ref{cgxi}), that the statistical moments of the coarse-grained field $\xi_{\hbox{cg}} (\mathcal{D}_\ell)$ scale across the inertial range as \cite{apol_etal},
\be
\mathbb{E}[ \xi_{\hbox{cg}}^q (\mathcal{D}_\ell) ]  \sim \ell^{\zeta_q} \ , \ \label{cgxiq}
\ee
with scaling exponent
\be
\zeta_q = \frac{\mu}{8}q(1-q) \label{zetaq} \ . \ 
\ee}}
In order to investigate the validity of (\ref{cgxiq}) and (\ref{zetaq}) it is necessary to record fluctuations of $N_\ell$, the number of vortex structures contained in $\mathcal{D}_\ell$. In fact, if we define
\be
M_q (\ell) \equiv \mathbb{E}[ N^q_\ell ] \label{moments1} \ , \ 
\ee
then $\zeta_q$ can be putatively obtained from
\be
\ell^{-2q} M_q (\ell) \sim \mathbb{E}[ \xi_{\hbox{cg}}^q (\mathcal{D}_\ell) ]  \sim \ell^{\zeta_q} \label{moments2} \ . \ 
\ee
A serious difficulty with the implementation of (\ref{moments1}) and (\ref{moments2}) is that vortices are not unambiguously defined physical objects, so that identifying and counting them can be a puzzling job. Among the several alternative ways of defining vortex structures \cite{hugo_mori}, we adopt the swirling strength criterion \cite{zhou_etal}, taking into account its simplicity (it is based on first order derivatives of the velocity field) and broad well-documented usage. 

\begin{figure}[h]
\begin{center}
\includegraphics[width=0.48\textwidth]{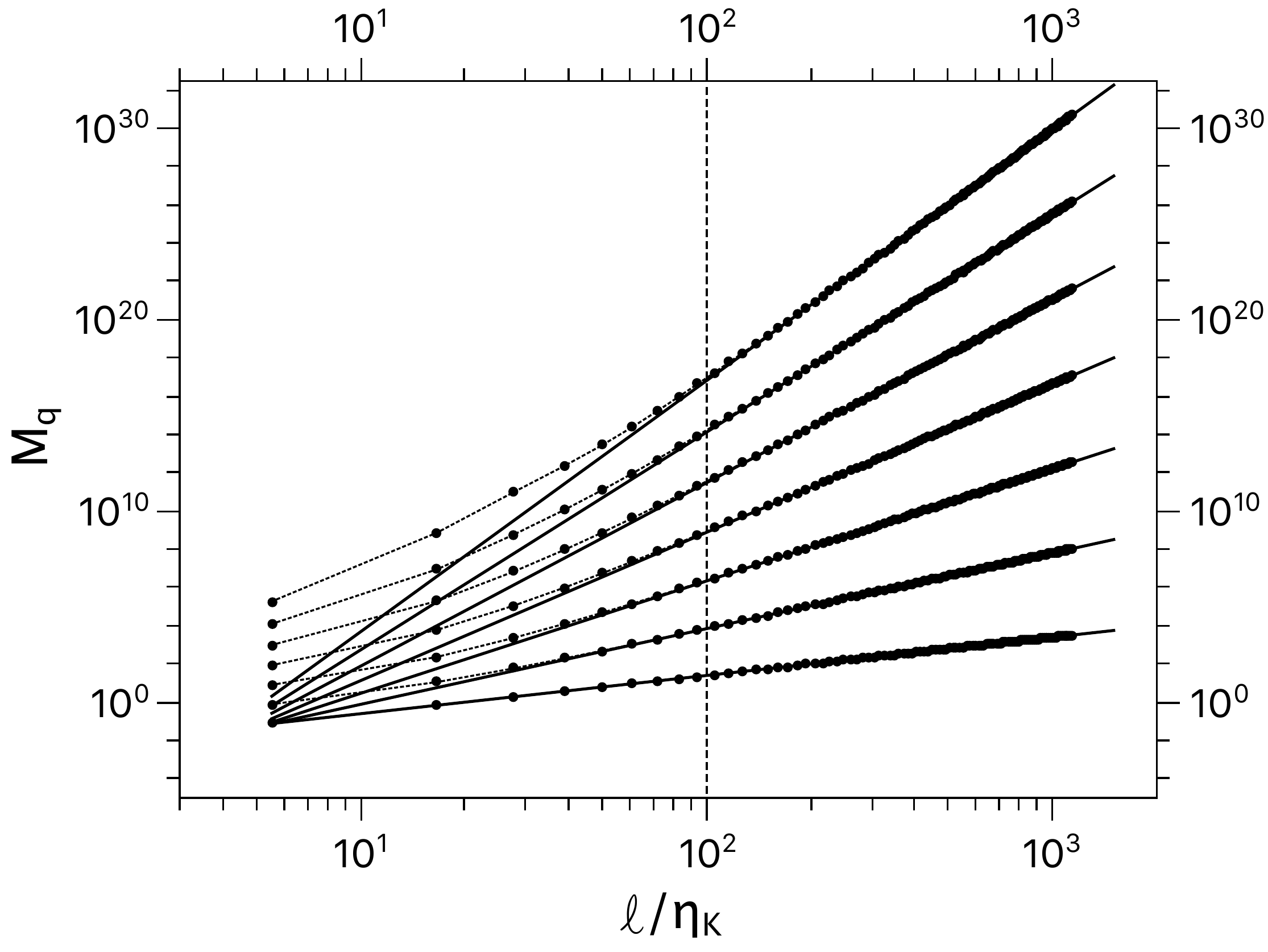}
\vspace{0.0cm}

\caption{Numerical evaluations of $M_q (\ell)$, Eq.~(\ref{moments1}), (circles connected by dotted lines) for $R_\lambda = 610$ and the related power law fits (solid lines) obtained in the scaling region {\hbox{{\textcolor{black}{$100 < \ell/\eta_K < 1130$}}}} (points at the right of the vertical dashed line). Graphs are vertically displaced to ease visualization and run through integer moment orders from $q=1$ at the bottom to $q=7$ at the top.}
\label{}
\end{center}
\end{figure}

\begin{figure}[h]
\begin{center}
\includegraphics[width=0.48\textwidth]{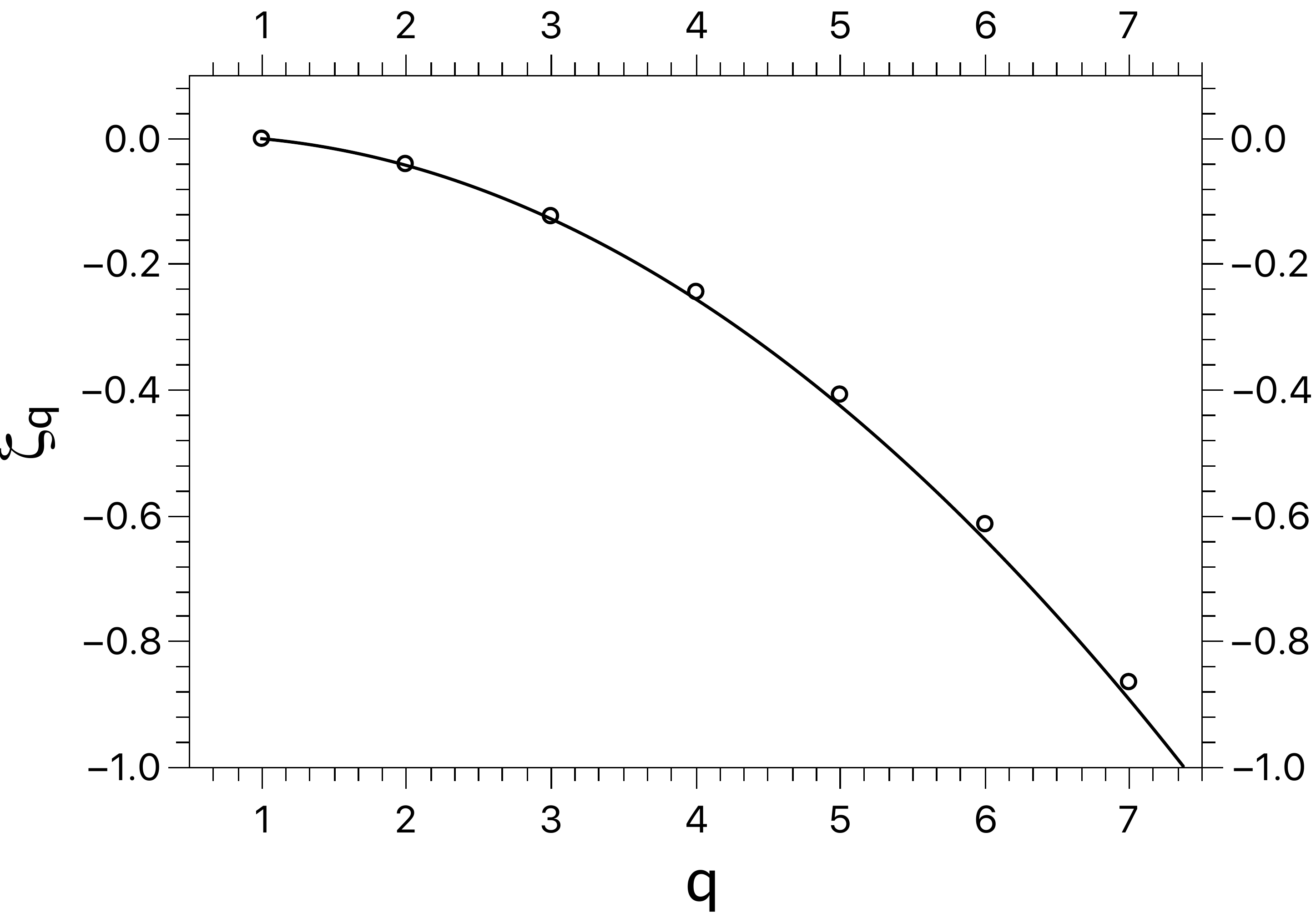}
\vspace{0.0cm}
\caption{The scaling exponents $\zeta_q$ for moment orders $q=1$ to 7 (open circles) obtained from the fits of Fig.~2 are compared to the predicted values given by Eq.~(\ref{zetaq}), with $\mu = 0.17$ (solid line). 
}
\label{}
\end{center}
\end{figure}

We have applied the swirling strength criterion to the velocity field projected on $4096^2$ planar slices of a $4096^3$ DNS domain ($R_\lambda = 610$) available from the Johns Hopkins turbulence database. Following the usual practice to avoid spurious results (fake structures) and poor vortex identification resolution (artificial vortex merging), vortices are identified whenever swirling strength eigenvalues have the absolute values of their imaginary parts larger than some prescribed threshold. In our specific case, we have identified vortices from the condition that {\hbox{$|$Im($\lambda$)$|$ $\geq$ 0.125$|$Im($\lambda$)$|_{\hbox{rms}}$}}. No relevant variations in the number of vortex structures are observed if the threshold factor 0.125 is changed to 0.5. This indicates that vortex structures form a well-resolved (dilute) system and are reasonably well captured by the swirling strength criterion.

From the example of Fig.~1, we see that the detected vortex structures have many different shapes. By inspecting the planar vorticity field within vortex spots, we assign to each of them a single representative point as the position where the absolute vorticity is maximum. A two-dimensional stochastic point process is then defined in this way. As described below, a large statistical ensemble with random values of $N _\ell$ is generated by sliding the square $\mathcal{D}_\ell$, site by site along the $4096^2$ sliced planar domains embedded in the host DNS grid.

Spanning the entire DNS spatial domain, we have worked with 64 evenly spaced two-dimensional slices normal to each of the three space directions, comprising a total number of 192 slices. To efficiently compute $N_\ell$ on the DNS slices, a function $f(\bf{r})$ is defined on the grid as unity at vortex representative positions and zero elsewhere. Then, $f({\bf r})$ is convolved (through a product in Fourier space) with an indicator function,
\be
\mathcal{I}_\ell(\bfr) = \left \{\begin{matrix}
{\hbox{1, if $|\bfr| \in {\mathcal{D}}_0$ ,}} \\
{\hbox{0, if $|\bfr| \notin {\mathcal{D}}_0$ ,}}
\end{matrix} \right.
\ee
where ${\mathcal{D}}_0$ is a square of side $\ell$ centered at the origin. The resulting function, $h({\bf r}) = \sum_{\bfr'} \mathcal{I}_\ell(\bfr -\bfr') f(\bfr')$, gives the total number of points inside a square of side $\ell$ centered at any arbitrary position ${\bf r}$ on the grid, which amounts to a realization of the two-dimensional random process $N_\ell$.

Results for the statistical moments of $N_\ell$ and the associated scaling exponents $\zeta_q$ are presented in Figs.~2 and 3. They provide excellent support to the postulated relationship between the surface vortex number density in planar slices of the flow and viscous dissipation, Eq.~(\ref{xi1}). We see in this way that the structural and multiplicative cascade aspects of turbulent intermittency are indeed closely connected. The local distribution of coherent vortex structures is found to reflect, at Kolmogorov length scales, the ``granularized" nature of the energy dissipation field $\epsilon(\bfr)$.

\section{A Monte Carlo Look at the Circulation Equations}

We focus now on the conjectured statistical equivalence between Eqs.~(\ref{circ}) and (\ref{circ2}). An interesting way to address this issue is through their respective characteristic functions, viz.,
\be
Z_1(\chi) \equiv  {\mathbb{E}} [ 
e^{i \chi \int_\mathcal{D} d^2 \bfr \xi(\bfr) \tilde \Gamma(\bfr)}
]_{\xi,\tilde \Gamma}
\ee
and
\be
Z_2(\chi) \equiv  {\mathbb{E}} [ 
e^{i \chi \xi_{\hbox{\scriptsize{cg}}}(\mathcal{D}) \int_\mathcal{D} d^2 \bfr \tilde \Gamma(\bfr) }
]_{\xi,\tilde \Gamma} \ . \
\ee
Upon averaging the above expressions over the Gaussian fluctuations of $\tilde \Gamma (\bfr)$, we find, in more compact notation,
\be
Z_m(\chi) =  {\mathbb{E}} [ e^{- \frac{1}{2} \chi^2 {\cal{N}}_m^{-1} \psi_m} ]_{\psi_m} \ , \ \label{Zm}
\ee
with $m=1,2$ and
\be
\psi_m \equiv {\cal{N}}_m \int_\mathcal{D} d^2 \bfr \int_\mathcal{D} d^2 \bfr' \xi(\bfr) \Delta_m(\bfr-\bfr') \xi(\bfr') \ , \ \label{psi}
\ee
where
\bea
&&\Delta_1(\bfr-\bfr') = (|\bfr-\bfr'| + \eta)^{-\alpha} \ , \ \label{delta1} \\
&&\Delta_2(\bfr-\bfr') = 1 \ , \ \label{delta2}
\eea
and the length scale $\eta$ plays the role of an ultraviolet regularization parameter (it can be identified to $\eta_K$) and ${\cal{N}}_m$ is a normalization constant adjusted so that the random variable $\psi_m$ gets described by a standardized probability distribution function with unit variance.


The characteristic function $Z_m(\chi)$, Eq.~(\ref{Zm}) is, of course, determined by the probability distribution functions $\rho_m(\psi_m)$ of $\psi_m$, for the bounded (finite $\Phi_0$) and unbounded ($\Phi_0 = \infty$) scalar field models of $\xi(\bfr)$. Aiming at determinations of $\rho_m(\psi_m)$, we have performed Monte Carlo simulations of the field theoretical model defined by (\ref{action}) and (\ref{V}).

Following Ref.~\cite{mori1}, we develop our numerical experiments with a two-dimensional $100 \times 100$ lattice which has lattice parameter $\eta$ (taken, for convenience, as the unity of length) and the case of bounded Liouville measures prescribed by $\Phi_0 = 2.7$ \cite{comment}. We furthermore define $\gamma =1$ for the intermittency parameter introduced in (\ref{xi_gm}) and {\textcolor{black}{$\alpha = 1.3$ ($\simeq 4/3 - \mu/2$)}} for the scaling exponent of the two-point correlation function (\ref{delta1}). 

\begin{figure}[t]
\includegraphics[width=0.48\textwidth]{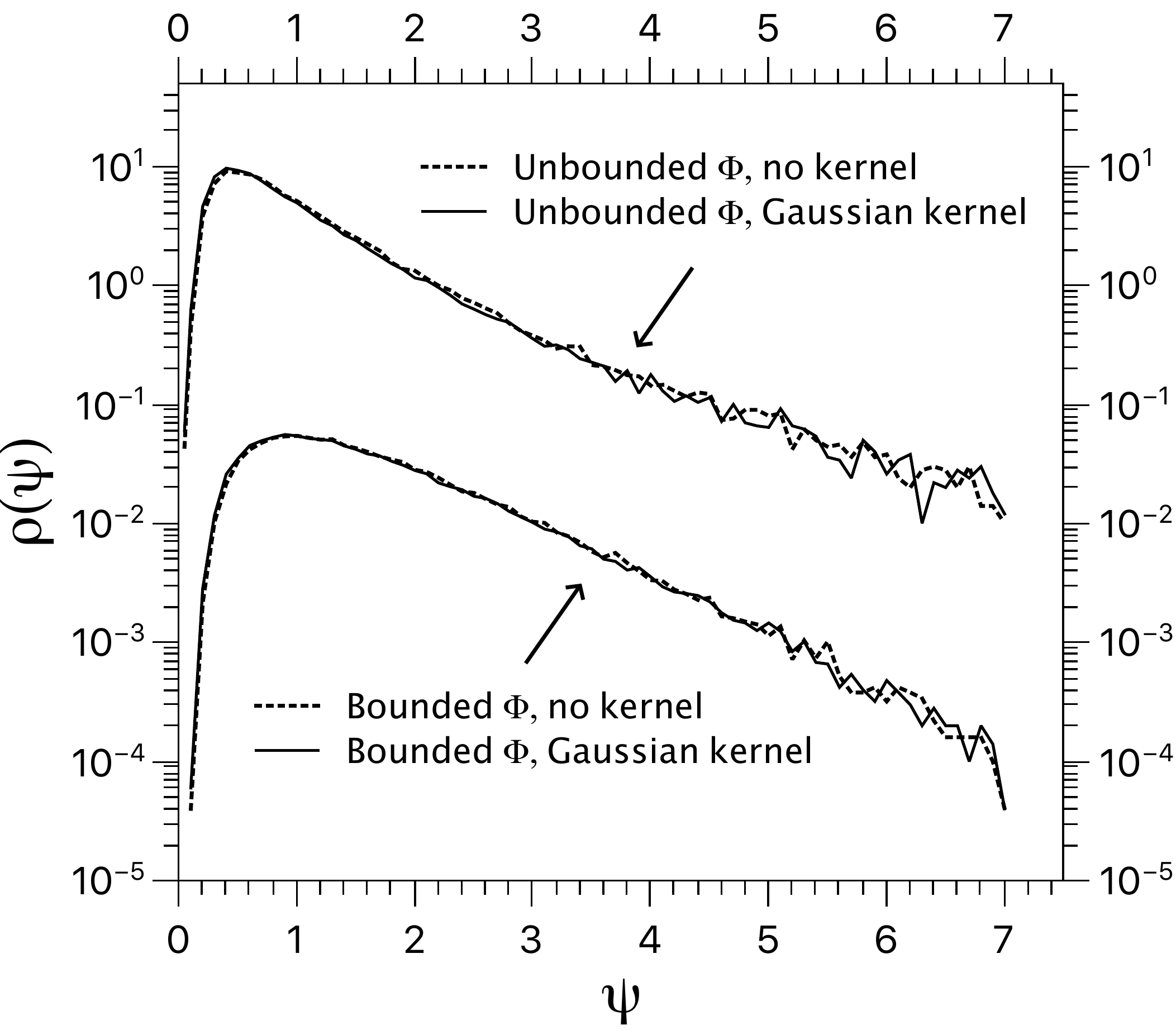}
\put(-190,145){{\large{(a)}}}

\includegraphics[width=0.48\textwidth]{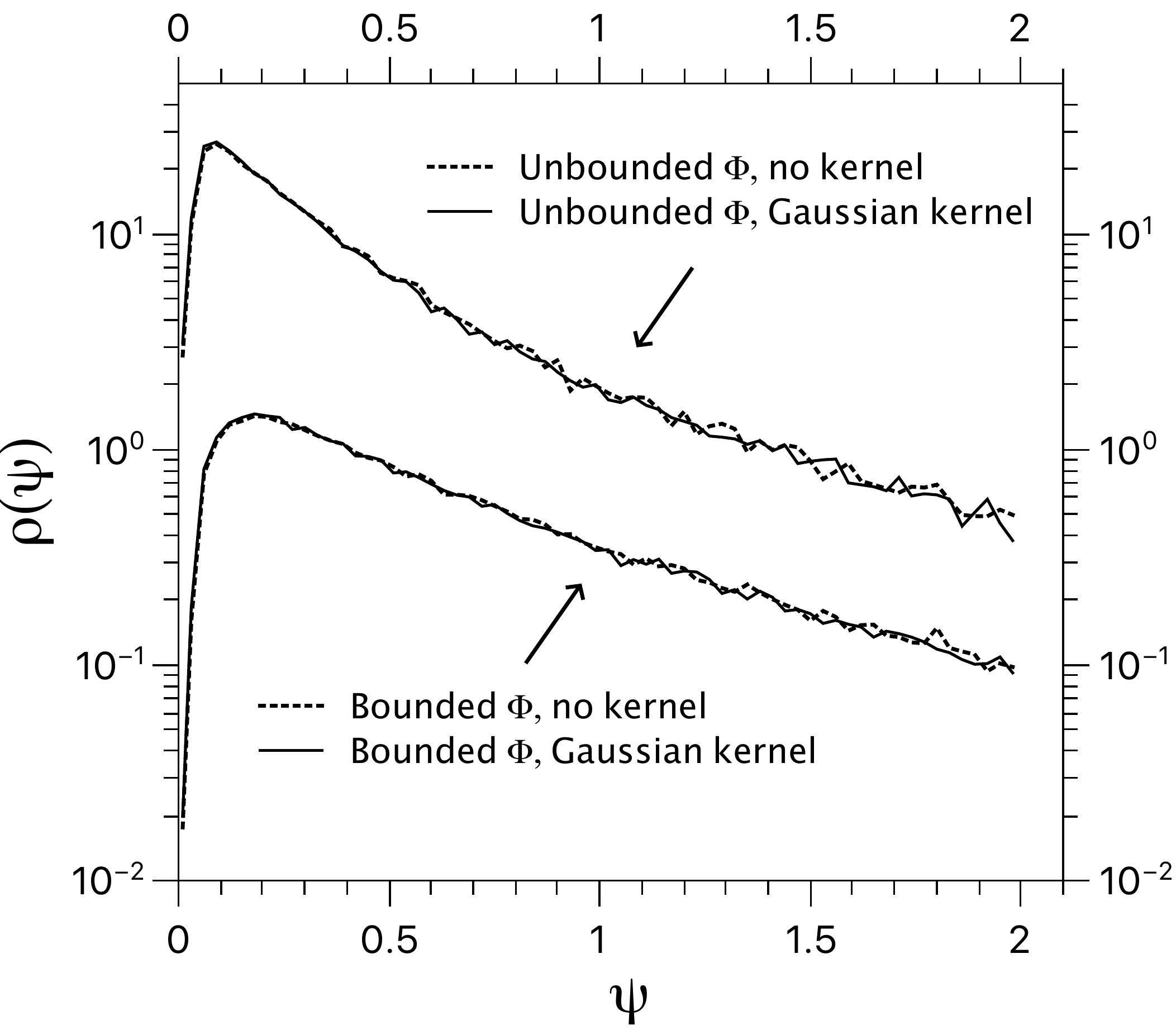}
\put(-190,145){{\large{(b)}}}
\vspace{0.0cm}
\caption{PDFs of the quadratic form $\psi$ for domains $\mathcal{D}$ of dimensions (a) $20 \eta \times 20 \eta$ and (b) $5 \eta \times 5 \eta$. Pairs of PDFs associated to the same lattice sizes have been vertically displaced to ease visualization; {\it{Bounded}} and {\it{unbounded}} $\Phi$ refer, respectively, to fluctuating fields {\hbox{$\phi \leq \Phi_0 = 2.7$, and $\phi < \infty$}}. The {\it{Gaussian kernel}} and {\it{no kernel}} captions indicate, respectively, the cases defined by Eqs.~(\ref{delta1}) and (\ref{delta2}).}
\end{figure}

The results reported in Fig.~4 fully corroborate our expectations. We study $\psi_m$ for blocks with two different dimensions, namely, $5 \eta \times 5 \eta$ (block in the scaling range) and $20 \eta \times 20 \eta$ (block in the integral scale). Fluctuations of the fine and the coarse-grained circulations, Eqs.~(\ref{circ}) and (\ref{circ2}), are observed not to depend on the specific details of the self-similar Gaussian kernels of the microcirculation field $\tilde \Gamma(\bfr)$, when cPDFs are recast in standardized form. In other words, {\hbox{$\rho_1(\psi) = \rho_2(\psi) \equiv \rho(\psi)$}}. It should be noted, however, as it can also be inferred from Figs.~4a and 4b, that $\rho(\psi)$ is strongly sensitive to the existence of a bound $\Phi_0$ for the fluctuations of the scalar field $\phi(\bfr)$. This fact is related to important phenomenological aspects of the turbulent circulation, as discussed in the following.
\vspace{-0.0cm}
\vspace{-0.5cm}



{\textcolor{black}{\section{Gaussian Factorization of Extreme Circulation Events}}}

We are mostly interested to understand the impact that the modified GMC model outlined in Sec.~II has on the structure of cPDFs. One may wonder, in this connection, whether the coarse-grained variable $\xi_{\hbox{cg}}$, Eq.~(\ref{cgxi}), is still lognormally distributed. As suggested by its low-order statistical moments, not much is changed, since the scaling exponents $\zeta_q$ are noticed to be well-approximated by quadratic functions of $q$ \cite{mori1}. This is reasonable, since low-order moments are dominated by relatively small fluctuations of $\phi (\bfr)$, which are not strongly perturbed by the {\textcolor{black}{bounded fluctuations of}} $\Phi_0$. In contrast, scalar field fluctuations that determine the scaling behavior of high-order moments of $\xi_{\hbox{cg}}$ are more frequently blocked by the bound $\Phi_0$, so that the quadratic profile of $\zeta_q$ is unavoidably lost and replaced by a linear one at large enough $q$ \cite{mori1}.

Not only large fluctuations of $\phi(\bfr)$ are affected in the modified GMC framework, but we also expect their correlation functions to become finite-ranged. The argument is simple: having in mind fluctuations of the free scalar field in the absence of any bound, consider the total area $A_{\Phi_0} < L^2$ of the planar region defined by $\phi(\bfr) < \Phi_0$. It is not difficult to obtain the Gaussian-based estimate
\be
A_{\Phi_0} \simeq \frac{1}{2} L^2 \left [ 1 + {\hbox{erf}} \left ( \frac{\Phi_0}{\sqrt{2} \phi_{rms}(L)} \right ) \right ] \ , \ \label{APhi0}
\ee
where \cite{mori1}
\be
\phi^2_{rms}(L) \equiv {\mathbb{E}}[\phi^2] = \frac{1}{2 \pi} 
\ln \left ( \frac{L}{\eta} \right ) \ . \ 
\ee
Eq.~(\ref{APhi0}) correctly leads to {\hbox{$A_{\Phi_0} =0$, $L^2/2$, and $L^2$}} in the respective limits
{\hbox{$\Phi_0 \rightarrow - \infty$, $0$, and $\infty$}}. A mass scale $m$ (inverse correlation length) is then introduced as
\be
m \sim \frac{1}{\sqrt{A_{\Phi_0}}} \ . \
\ee
All of the above means, in the same fashion, that fluctuations of $\xi(\bfr)$ are {\textcolor{black}{bounded from above}} and finite-ranged in the modified GMC model. We may suspect, thus, on the grounds of the central limit theorem, 
that the random variable $\xi_{\hbox{cg}}(\mathcal{D})$ crosses over, at fixed $\ell$, from lognormal 
to Gaussian behavior as $\Phi_0$ gets smaller. We have investigated this phenomenon through Monte-Carlo simulations which have the same defining parameters $L/\eta = 100$ and $\gamma = 1$ as the ones introduced in Sec.~IV. Fig.~5 yields results for domains with sizes $\ell/\eta = 50$ (Fig.~5a) and $\ell/\eta = 5$ (Fig.~5b), for a sample of bounds $\Phi_0$.

\begin{figure}[t]
\includegraphics[width=0.48\textwidth]{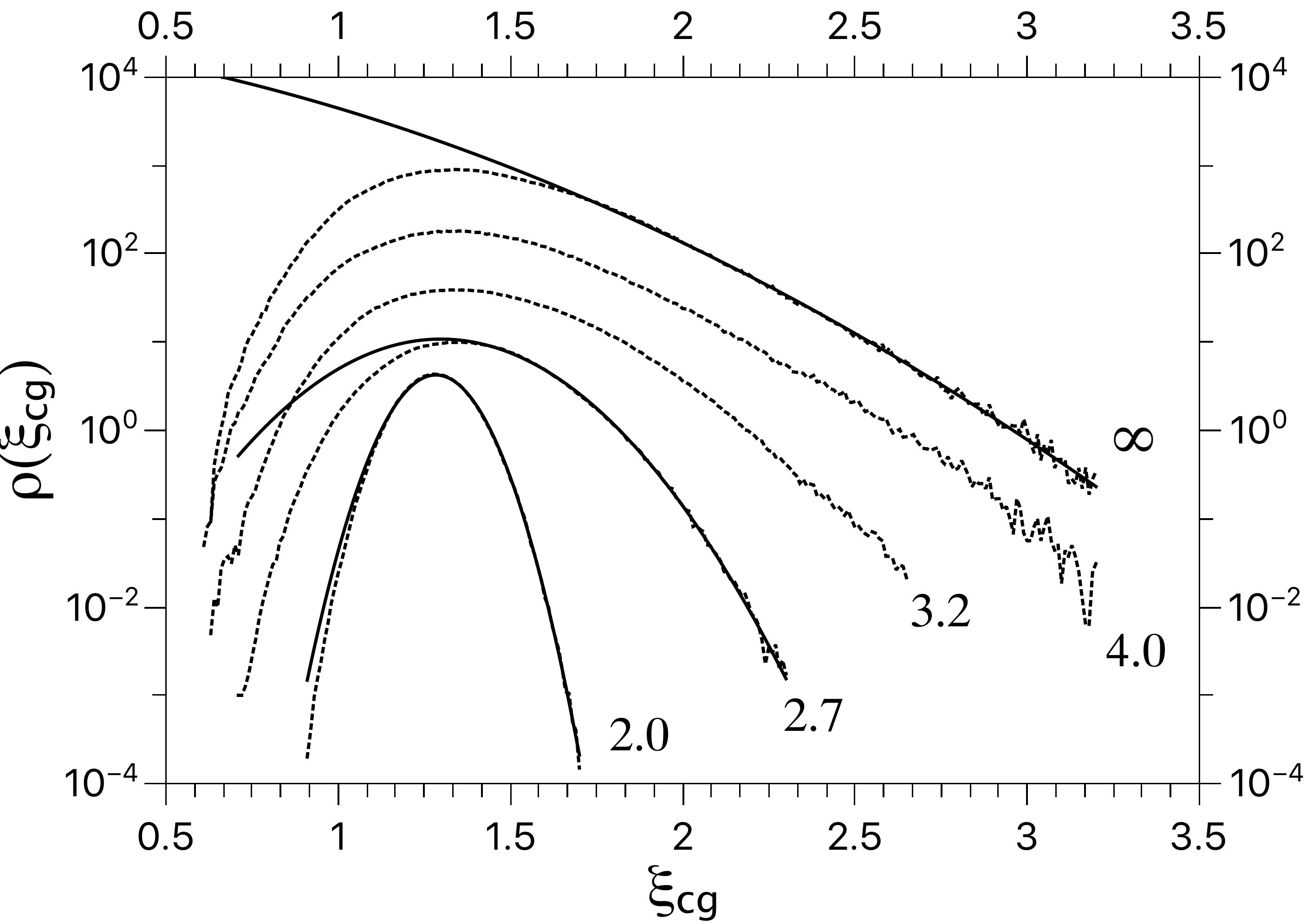}
\put(-120,140){{\large{(a)}}}

\includegraphics[width=0.48\textwidth]{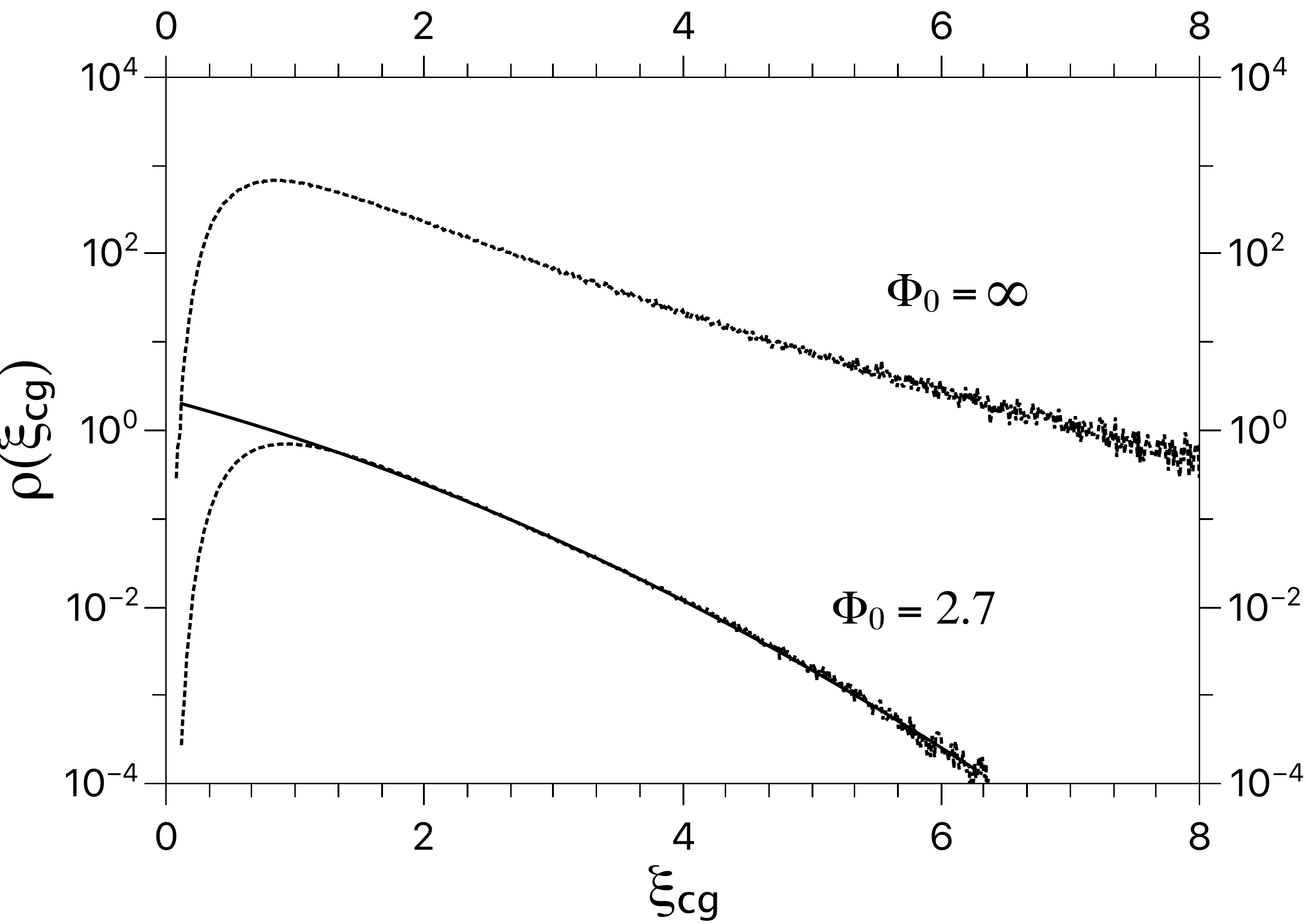}
\put(-120,140){{\large{(b)}}}
\vspace{0.0cm}
\caption{PDFs of the coarse-grained field $\xi_{\hbox{cg}}(\mathcal{D})$ (dashed lines) for unbounded 
{\textcolor{black}{$(\Phi = \infty)$ and bounded}} fluctuations of $\phi$ defined by
$\Phi_0 =2.0, 2.7,3.2,$ and $4.0$. Solid lines show the parabolic interpolations (conjectured Gaussian approximations) for some of the concave PDFs' right tails. Coarse-graining is carried out for domains $\mathcal{D}$ of dimensions (a) $50 \eta \times 50 \eta$ and (b) $5 \eta \times 5 \eta$.}

\label{pdfs}
\end{figure}

\begin{figure}[t]
\begin{center}
\includegraphics[width=0.48\textwidth]{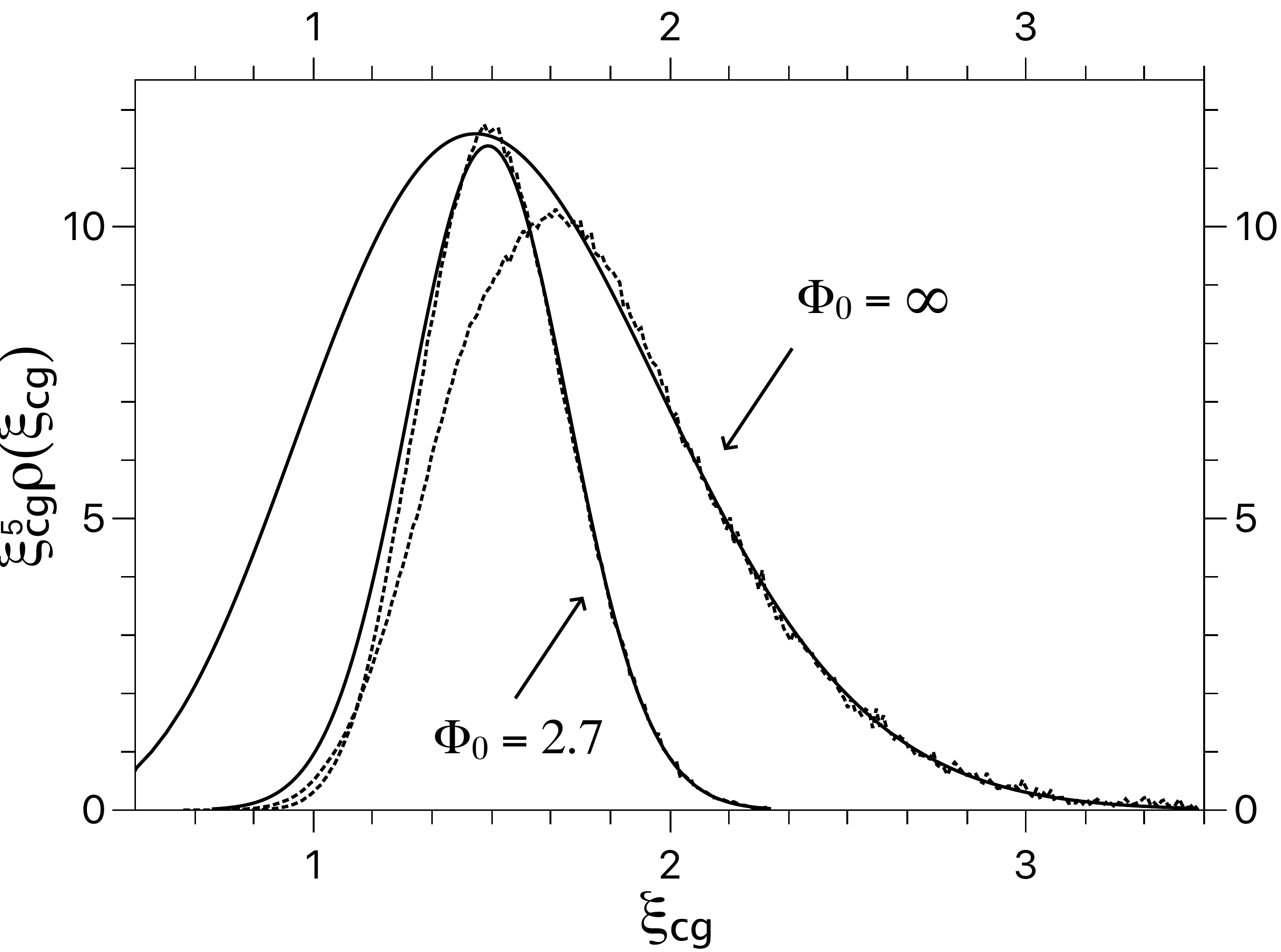}
\put(-200,140){{\large{(a)}}}

\includegraphics[width=0.48\textwidth]{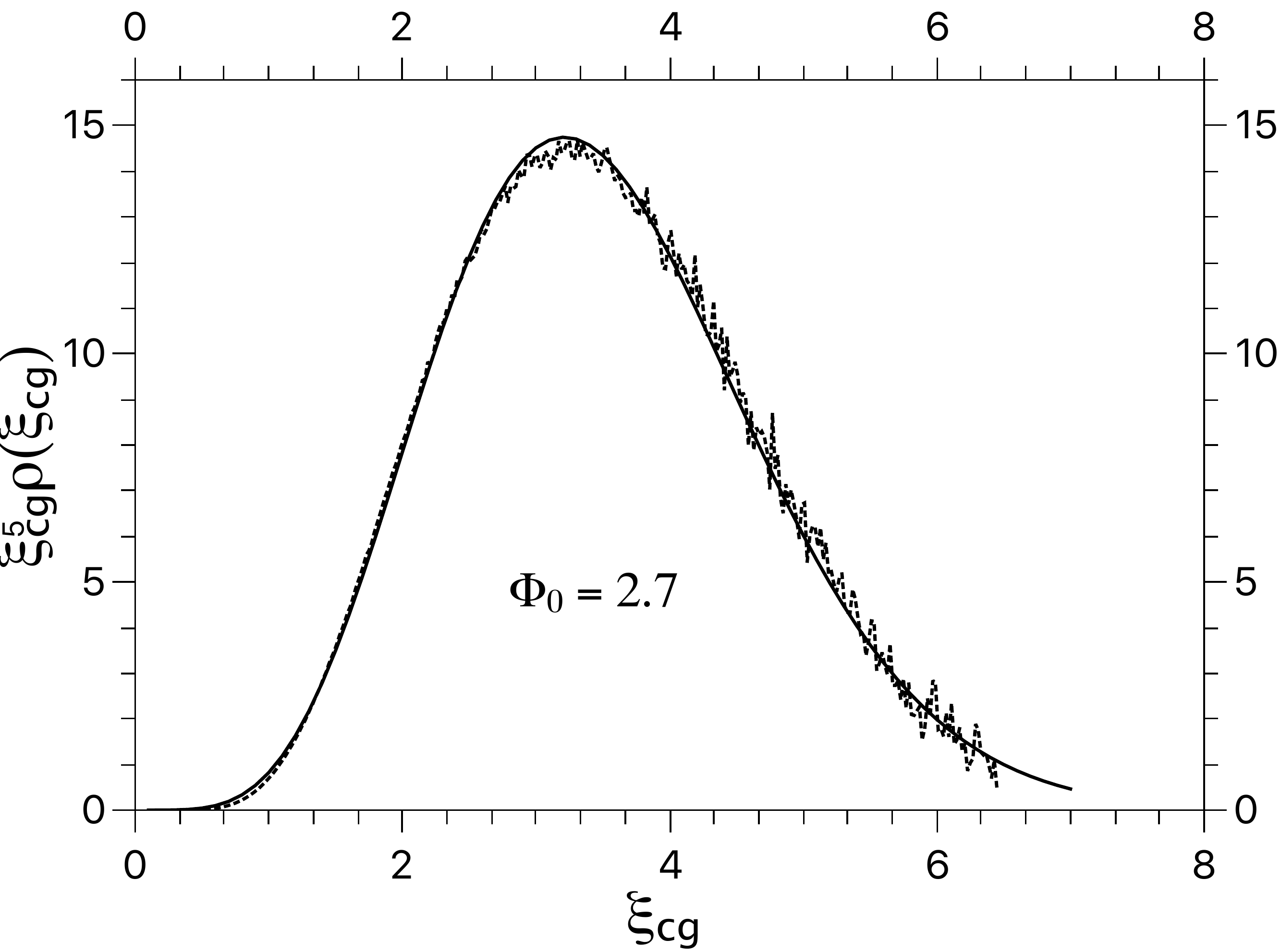}
\put(-200,140){{\large{(b)}}}
\vspace{0.0cm}
\caption{Dashed and solid lines correspond, respectively, to the dashed and solid lines 
of Fig.~5. Panels show the integrands of the fifth-order moments of $\xi_{\hbox{cg}}(\mathcal{D})$ for 
domains $\mathcal{D}$ of dimensions (a) $50 \eta \times 50 \eta$ and (b) $5 \eta \times 5 \eta$.} 
\label{pdfs}
\end{center}
\end{figure}

As it can be noticed for the case $\ell/\eta = 50$ in {\hbox{Fig.~5a}}, $\xi_{\hbox{cg}}(\mathcal{D})$ follows more closely a normal distribution the smaller is $\Phi_0$. 
However, once we want to discuss extreme fluctuations of the circulation variable (\ref{circ2}), an accurate Gaussian approximation for the whole PDF of $\xi_{\hbox{cg}}(\mathcal{D})$ may be a superfluous, too restrictive, condition. A relevant point in this context is to check whether large positive fluctuations of $\xi_{\hbox{cg}}(\mathcal{D})$ can be well described by Gaussian PDF tails (the solid lines in Figs.~5a and 5b), and to what extent such relaxed approximations lead to reliable evaluations of high-order moments of $\xi_{\hbox{cg}}(\mathcal{D})$. 

Special care is needed here not to be fooled by appearances. In fact, even though the PDF's right tail of $\xi_{\hbox{cg}}(\mathcal{D})$ for the unbounded case ($\Phi_0 = \infty$) seems to be reasonably well approximated by a Gaussian tail, as indicated in Fig.~5a, it fails to account for the evaluation of the fifth-order moment of $\xi_{\hbox{cg}}(\mathcal{D})$, see Fig.~6a. In opposition, much better results for the same moment order are achieved for the two studied sizes of $\mathcal{D}$ in the situation where $\Phi_0=2.7$, according to the related plots provided in Figs.~6a and 6b.

A more careful analysis of the pertinence (or not) of a Gaussian description of the large positive fluctuations of $\xi_{\hbox{cg}}(\mathcal{D})$ can be put forward with the help of a conveniently defined set of expectation values, as detailed below.
\vspace{0.2cm}

\noindent $\bullet$ $\mathbb{E}[\xi_{\hbox{cg}}^q(\mathcal{D})]$ $\equiv$
statistical moment determined directly from the Monte Carlo samples of
$\xi_{\hbox{cg}}(\mathcal{D})$;
\vspace{0.2cm}

\noindent $\bullet$ $\mathbb{E_G}[\xi_{\hbox{cg}}^q(\mathcal{D})]$ $\equiv$  statistical moment determined by the (not normalized) Gaussian distributions that fit the right tails of the PDFs of $\xi_{\hbox{cg}}(\mathcal{D})$;
\vspace{0.2cm}

\noindent $\bullet$ $\mathbb{E_{LG}}[\xi_{\hbox{cg}}^q(\mathcal{D})]$ $\equiv$ statistical moment determined by a lognormal distribution which has its mean and variance computed from the Monte Carlo samples of $\ln[\xi_{\hbox{cg}}(\mathcal{D})]$.
\vspace{0.2cm}

\begin{figure}[h]
\begin{center}
\vspace{0.0cm}
\includegraphics[width=0.48\textwidth]{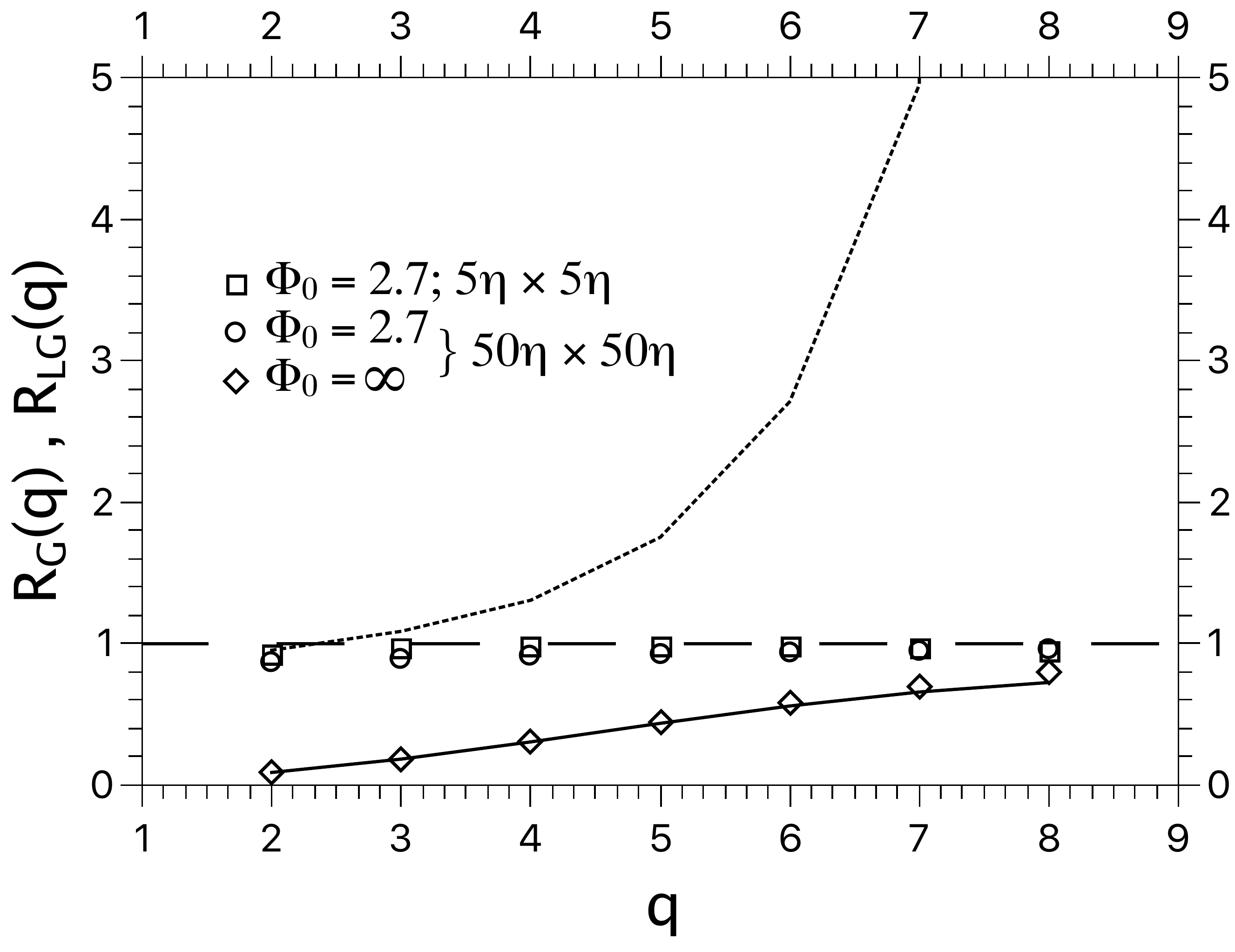}
\vspace{0.0cm}
\caption{Statistical moment ratios $R_\mathbb{G}(q)$ (symbols) and 
$R_\mathbb{LG}(q)$ (short-dashed line for $\Phi_0 = 2.7$ 
and domain $\mathcal{D}$ of dimensions $5 \eta \times 5 \eta$; 
solid line for $\Phi_0 = \infty$ and domain $\mathcal{D}$ of dimensions $50 \eta \times 50 \eta$).}
\label{}
\end{center}
\end{figure}

\noindent We introduce, accordingly, a pair of statistical moment ratios $R_\mathbb{G}(q)$
and $R_\mathbb{G}(q)$,

\be
R_\mathbb{G}(q) \equiv \frac{\mathbb{E}[\xi_{\hbox{cg}}^q]}{\mathbb{E_G}[\xi_{\hbox{cg}}^q]} \ , \ 
R_\mathbb{LG}(q)  \equiv \frac{\mathbb{E_{LG}}[\xi_{\hbox{cg}}^q]}{\mathbb{E_G}[\xi_{\hbox{cg}}^q]} 
\ . \
\ee
If, therefore, $R_\mathbb{G}(q) \simeq 1$ holds for high moment orders, then large positive deviations of $\xi_{\hbox{cg}}(\mathcal{D})$ behave like Gaussian random variables. On the other hand, the scenario of Gaussian fluctuations loses its relevance if $R_\mathbb{G}(q) \simeq R_\mathbb{LG}(q)$, once the lognormal description of $\xi_{\hbox{cg}}(\mathcal{D})$ is found to be as effective as in the pure (original) GMC model. 

Fig.~7 shows representative evaluations of $R_\mathbb{G}(q)$ and $R_\mathbb{LG}(q)$
at integer orders $2 \leq q \leq 8$. The $\Phi_0 = \infty$ case for $\ell / \eta = 50$ is found to be 
excellently accounted for by a lognormal description, something that does not happen for the bounded cases 
given by $\Phi_0 = 2.7$. 

The essential message conveyed from the investigated statistical moment ratios $R_\mathbb{G}(q)$ and $R_\mathbb{LG}(q)$ is that the modified GMC model yields PDFs of $\xi_{\hbox{cg}}(\mathcal{D})$ which are characterized by Gaussian right tails. Under such circumstances, we may rewrite $\xi_{\hbox{cg}}(\mathcal{D})$ as
\be
\xi_{\hbox{cg}}(\mathcal{D}) = \mathbb{E}[\xi_{\hbox{cg}}(\mathcal{D})] + \tilde \xi_{\hbox{cg}}(\mathcal{D}) 
\ , \ \label{split}
\ee
where $\tilde \xi_{\hbox{cg}}(\mathcal{D})$ has vanishing mean and Gaussian-like fluctuations for 
$\tilde \xi_{\hbox{cg}}(\mathcal{D}) \gg \mathbb{E}[\xi_{\hbox{cg}}(\mathcal{D})]$.

Substituting, now, (\ref{split}) into (\ref{circ2}), we get, for the circulation variable,
\be
\Gamma = \mathbb{E}[\xi_{\hbox{cg}}(\mathcal{D})] \int_\mathcal{D} d^2 \bfr \tilde \Gamma(\bfr) 
+  \tilde \xi_{\hbox{cg}}(\mathcal{D}) \int_\mathcal{D} d^2 \bfr \tilde \Gamma(\bfr)
\ . \ \label{circ3}
\ee
The two contributions on the right-hand side of Eq.~(\ref{circ3}) have completely different large deviation behaviors. The first one is clearly Gaussian; the second is given as the product of two Gaussian random variables -- a well-known case study in probability theory \cite{craig} -- and turns out to dominate the positive or negative large deviations of $\Gamma$. These are then predicted to follow cPDF tails which have the same asymptotic decay as modified Bessel's functions of the second kind, that is,
\be
\rho(\Gamma) \sim \exp \left ( - c |\Gamma| \right )/\sqrt{|\Gamma|} \ , \ \label{cPDFtails}
\ee
for some positive constant $c$. {\textcolor{black}{There is an}} impressive agreement between (\ref{cPDFtails}) {\textcolor{black}{and the functional form of $\rho(\Gamma)$ predicted through the instanton approach \cite{migdal2020}. Circulation instantons leading to (\ref{cPDFtails}) were proposed by Migdal as velocity configurations that dominate the characteristic function of the circulation variable. Statistical averages are, in that setting, evaluated from a random ensemble of generalized Beltrami flows, conjectured to dissipate energy at small viscous scales. It is worth stressing, furthermore, that Eq.~(\ref{cPDFtails}) is strongly supported by high Reynolds number numerical simulations \cite{Iyer_etal_PNAS}.}}
\vspace{0.5cm}

\section{Conclusions}
\vspace{-0.0cm}

We have carefully examined important conceptual and technical assumptions of the vortex gas model of turbulent circulation statistics \cite{apol_etal} in its (improved) modified GMC version \cite{mori1}. They provide the stage for our derivation, as a central phenomenological result, of the decaying form of cPDF tails, the asymptotic relation (\ref{cPDFtails}). We find a perfect (and somewhat surprising) analytical agreement with the description of circulation intermittency based on the probability of occurrence of statistically dominant (instanton) configurations of the velocity field \cite{migdal2020}.
\vspace{-0.0cm}

{\textcolor{black}{This works provides}} strong support for the proposed relation between the dissipation field $\epsilon(\bfr)$ and the distribution of turbulent vortex structures in planar slices of the flow, {\textcolor{black}{a phenomenological connection which is also observed in the distribution of quantum vortices in superfluid turbulence \cite{muller_etal}}}. The interpretation of {\hbox{$\xi(\bfr) \propto \sqrt{\epsilon(\bfr)}$}} as the surface number density of planar vortex structures is quantitatively addressed in Sec.~III with striking success. This is clearly evidenced from the comparison of the predicted parabolic profile of the scaling exponents $\zeta_q$ for the $q^{th}$-order statistical moments of {\textcolor{black}{$\xi_{\hbox{cg}}(\mathcal{D}_\ell)$}} and their empirical evaluations, as shown in Fig.~3.

{\textcolor{black}{As an interesting motivating point for future research, we address the conjecture that the upper boundedness of $\xi(\bfr)$ is a consequence of the existence of repulsive statistical interactions between vortex structures at small dissipative scales \cite{mori_pereira_valadao}. In fact, mutual vortex repulsion is likely to preclude high density fluctuations of planar vortex clusters. Along these lines, we note that precise numerical estimates of the bounding parameter $\Phi_0$, introduced in Eq.~(\ref{V}), are still in order.}}

{\textcolor{black}{We call attention, furthermore, to the case of non-planar circulation contours, an essentially open problem, where minimal surfaces \cite{migdal2020, Iyer_etal_PNAS} are expected to play a vital role in the statistical description of turbulent circulation fluctuations. Understanding the need for minimal surfaces remains as a major challenge within the present vortex gas model of circulation statistics, even though the application of Eq.~(\ref{circ}) to arbitrary bounded surfaces is in principle free of modeling difficulties.}}
\vspace{0.5cm}

\leftline{{\it{Acknowledgements}}}
\vspace{0.3cm}

The authors thank V.J. Valadão for fruitful discussions.








\vspace{0.0cm}


\begin{references}
\bibitem{comment} The specific choice $\Phi_0 = 2.7$ retaken from Ref.~\cite{mori1} is not intended to model empirical (DNS) data (which is out of the scope of this work), although it is probably not far from a realistic estimate. We are here rather interested to study some general features of the modified GMC model.
\end{references}
\end{document}